\documentclass[prd,aps,twocolumn]{revtex4}

\usepackage[dvips]{graphicx}

\def\be{\begin{equation}}
\def\ee{\end{equation}}
\def\bi{\begin{itemize}} 
\def\ei{\end{itemize}}
\def\ben{\begin{enumerate}}
\def\een{\end{enumerate}}

\begin{document}

\title{Constraint likelihood analysis for a network of gravitational wave detectors}

\author{S.~Klimenko, S.~Mohanty\dag, M.~Rakhmanov and G.~Mitselmakher}

\affiliation{University of Florida, P.O.Box 118440, Gainesville, Florida, 32611, USA}
\affiliation{\dag The University of Texas at Brownsville, 80 Fort Brown, 
Brownsville, Texas, 78520, USA}

\begin{abstract}
We propose a coherent method for the detection and reconstruction of
gravitational wave signals with a network of interferometric 
detectors. 
The method is derived using the likelihood functional
for unknown signal waveforms. 
In the standard approach, the global
maximum of the likelihood over the space of waveforms is used as the detection
statistic.  
We identify a problem with  this approach.  
In the case of an aligned pair of detectors, the detection statistic
depends on the cross-correlation 
between the detectors as expected, but this dependence dissappears even
for infinitesimally small misalignments. We solve
the problem by applying constraints on the
likelihood functional  and 
obtain a new class of statistics. The resulting method can be applied to
data from a network consisting of any number of detectors with
arbitrary detector orientations.   
The method allows us reconstruction of the source coordinates and the waveforms 
of two polarization components of a gravitational wave.
We study the performance of the method with numerical simulation and find
the reconstruction of the source coordinates to be more accurate 
than in the standard approach.
\end{abstract}

\maketitle


\section{Introduction}

Several  gravitational wave (GW) detectors are now operating around the world, 
including both laser interferometers 
\cite{LIGO, VIRGO, GEO600, TAMA300} and resonant mass 
detectors \cite{IGEC}. Combining the data from such a 
network of detectors can benefit both the detection of  GW signals and estimation 
of signal parameters. Unlike real GW signals that would occur 
in coincidence across all detectors in a network,
most background events due to instrumental and terrestrial disturbances are 
expected to be local to each detector and, therefore, can be rejected by  
analysing data from a network. 
Given a network with different orientations and locations of the detectors, 
GW sources can be localized on the sky and the waveforms of the two 
independent GW polarization components can be reconstructed.

Methods for the analysis of data from a network of GW detectors 
can be divided into two classes: {\em coincidence} and 
{\em coherent} methods.
In coincidence methods, first, a search for GW signals is carried out for 
individual detectors and a list of candidate events is generated. 
Then a subset of events is selected by requiring temporal
coincidence of events between the detectors.
In coherent methods, one, first, combines the detector responses 
and then analyzes the combined data to generate a single list of
events. 

Networks of detectors are particularly important for searches
of gravitational wave {\em burst} signals. These are defined to be broadband
signals that may come either from unanticipated sources
or from sources  for which no reliable theoretical prediction exists for signal 
waveforms. Potential astrophysical sources of burst 
signals are stellar core collapse in Supernovae~\cite{corecollapse},
mergers of binary neutron star or black hole 
systems~\cite{flanagan+hughes:I} and
Gamma Ray Burst progenitors~\cite{Kobayashi+Meszaros}.

The first coherent method for burst searches with a network of three 
misaligned detectors was proposed by G\"{u}rsel and Tinto~\cite{gursel+tinto}.
In this method, the detector responses are combined into 
a functional, which attains its minimum at the correct direction 
to the source. The minimization of the functional allows one to  
reconstruct the source coordinates and two polarization waveforms 
of the burst signal.

Flannagan and Hughes~\cite{flanagan+hughes:II} considered maximization of the
 {\em likelihood functional}~\cite{helstrom,kendall}
 as a means of reconstructing source direction and polarization waveforms. 
 Anderson {\em et al}~\cite{epower} extended this approach 
to derive a detection statistic called {\em excess power}.
It is obtained by integrating the 
likelihood functional, weighted by a Bayesian prior probability density, 
over the space of all waveforms. In this paper,  
we refer to signal detection and reconstruction based on 
 the global maximum of the unweighted likelihood functional as 
the {\em standard likelihood} method.
Another coherent method, proposed by Sylvestre \cite{sylvestre}, 
starts with the {\em ad hoc} approach of forming a 
linear combination of data from a network of detectors. 
The combination coefficients are then
adjusted to construct a quadratic detection algorithm that satisfies certain well defined
criteria. 

Arnaud {\em et al}~\cite{arnaud:coherent_det} have numerically explored 
the issue of statistical performance of coherent and coincidence 
methods and find that the former  
are more efficient than the latter for burst signals.
In the case of signals with known 
waveforms, Finn~\cite{finn:2001} has shown using simulations that a coherent method
can also be robust when confronted with non-Gaussian noise. 

Coherent methods can also be used for rejecting coincident background 
signals. Cadonati~\cite{cadonati:rstat} has proposed 
a cross-correlation test, 
called {\em r-statistic}, for pairs of aligned detectors as a follow up 
consistency check on a coincidence analysis~\cite{LIGO_BURST_PAPER}. 
Rakhmanov and Klimenko~\cite{rakhmanov+klimenko} have proposed 
the {\em mixed correlations} method that extends the cross-correlation 
test to a network of three or more misaligned detectors.
Wen and Schutz~\cite{wen+schutz} have recently generalized the coherent
approach of G\"ursel and Tinto~\cite{gursel+tinto} to create a method for rejecting background coincident
 signals with a network of arbitrary detectors.

In this paper, we propose a method
for the coherent detection and reconstruction of burst signals that is based 
on the use of the likelihood ratio~\cite{helstrom,kendall}. 
Our analysis differs from~\cite{flanagan+hughes:II,epower} in an important way. 
We identify and solve a problem with the standard likelihood analysis, 
first spotted in ~\cite{Johnston:msthesis}.
The problem, which we call 
the {\em two detector paradox}, is that the maximum likelihood ratio
statistic for misaligned detectors
does not reduce, contrary to physical intuition, to the statistic 
for co-aligned detectors in the limit of
 small misalignment angles. The latter statistic depends on 
the cross-correlation of detector outputs whereas the former does not.
 We show that the problem originates in the maximization of 
the likelihood ratio functional
over all signal waveforms including those to which a detector network 
may not actually be sensitive. 
We propose a solution to this problem  that is based on 
constraints imposed on the GW signal waveforms. 

The constrained maximization of the likelihood functional
yields new detection and reconstruction methods which we call 
the {\em constraint likelihood} methods. 
Unlike the G\"{u}rsel and Tinto method, the constraint likelihood methods
can be used for arbitrary networks, including networks consisting of two detectors.
The performance of these methods  is studied in comparison with 
the standard likelihood method by using the numerical simulations.
In the simulation we use  networks of interferometric detectors consisting of 
LIGO 4~km detector in Hanford (H1), LIGO 4~km detector in Livingston (L1),
GEO-600 detector (G1), TAMA detector (T1) and VIRGO detector (V1). 
We find that the constraints employed in this paper
enhance the detection efficiency of the likelihood method. For detected sources,
the constraints significantly improve accuracy of the source 
localization.


The rest of the paper is organized as follows. Section~\ref{s_network}
lays out much of the basic notation and conventions used in the paper.
In Section~\ref{overview}, we provide an overview of the standard likelihood approach 
and its application to burst signals.  Section~\ref{2dp} describes 
the two detector paradox that appears in the standard likelihood approach. 
The origin of this problem is discussed in Section~\ref{nresp}. 
In  Section~\ref{constraint} we derive the constraint likelihood methods. 
The results from numerical studies of the performance of these methods 
are described in Section~\ref{Results}. 


\section{Detector response to gravitational waves}
\label{s_network}

\subsection{Gravitational wave signal}

Gravitational waves are described by a symmetric tensor of second rank 
$h_{ij}(t)$, which is usually defined in the transverse-traceless
gauge~\cite{Thorne:1987}. It takes particularly simple form in the 
coordinate frame associated with the wave. 
%
%
%
%
%
%
In this coordinate frame (the wave frame), a gravitational wave propagates in the
direction of $z$ axis and it can be described with the waveforms $h_{+}(t)$ and 
$h_{\times}(t)$ representing two independent polarizations components of the wave.

In addition to the waveforms $h_{+}(t)$ and $h_{\times}(t)$ 
we will use complex waveforms defined as
\begin{eqnarray}
   u(t)  =  h_{+}(t) + i h_{\times}(t) , \\
   \tilde{u}(t)  =  h_{+}(t) - i h_{\times}(t) .
\end{eqnarray}
In what follows tilde will always denote complex conjugation. 
The GW waveforms $u(t)$ and $\tilde{u}(t)$ are eigenstates of the rotations 
around $z$-axis in the wave frame. We denote this
particular rotation by $R_z(\psi)$, where $\psi$ is 
the rotation angle. 
The rotation $R_z(\psi)$ generates equivalent waveforms
which are different representations of the same gravitational wave.

We define the sum-square energy 
\footnote{This quantity is not the total physical energy carried by the 
gravitational wave.} 
carried by the gravitational wave as
\begin{equation}
\label{Ess}
   E = \int_{-\infty}^{\infty} \left( h^2_{+}(t)  + h^2_{\times}(t) \right) \; dt = 
      \int_{-\infty}^{\infty} u(t)  \tilde{u}(t) \; dt .
\end{equation}
Note that the sum-square energy is invariant under the rotation
$R_z$.

\subsection{Detector response}
\label{dresponse}

The response of the interferometer to an arbitrary gravitational wave 
$h_{ij}(t)$ is given by 
\begin{equation}
   \xi(t) = \frac{1}{2} \; T_{ij} \, h_{ij}(t) ,
\end{equation}
where $T_{ij}$ is the detector tensor~\cite{Dhurandhar:1988}.
In the wave frame the detector response is a linear superposition of two 
GW polarizations
\begin{equation}
   \xi(t) = F_{+} h_{+}(t) + F_{\times} h_{\times}(t) .
\end{equation}
where the coefficients $F_{+}$ and $F_{\times}$ are known as antenna patterns. 

To calculate the antenna pattern we introduce
the Earth-centered frame described in \cite{lazzarini}.
In this frame the detector location is defined by 
a radius-vector $\mathbf{r}$ pointing to the detector and its orientation 
is described by two unit vectors $\mathbf{a}$ and $\mathbf{b}$ along the 
detector arms. The vectors $\mathbf{a}$ and $\mathbf{b}$ define the detector tensor
\begin{equation}\label{defDetTensor}
   T'_{ij} = a_{i} \, a_{j} - b_{i} \, b_{j}, ~~i,j=1,2,3,
\end{equation}
where the indices correspond to spatial coordinates $x$, $y$ and $z$ respectively.
The direction to the GW source is defined in the Earth-centered frame
by two spherical angles $\phi$ (longitude) and $\theta$ (lattitude).
%
The rotational transformation
which connects the Earth-centered frame with the wave frame is given by
\begin{equation}
   {\mathbf{R}}(\phi, \theta) = {\mathbf{R}}_y(\theta) 
      {\mathbf{R}}_z(\phi) .
\end{equation}
It defines  the detector tensor in the wave frame 
\begin{equation}
   {\mathbf{T}}(\phi, \theta) = {\mathbf{R}}(\phi, \theta) \; 
       {\mathbf{T'}} \; {\mathbf{R}}(\phi, \theta)^T .
\end{equation}
Omitting the explicit dependence on the angles, the antenna patterns 
corresponding to the $h_+(t)$ and $h_{\times}(t)$ polarizations are calculated as follows:
\begin{eqnarray}
   F_{+}      & = & \frac{1}{2} 
      \left( T_{11} - 
             T_{22} \right) , \\
   F_{\times} & = & \frac{1}{2}
      \left( T_{12} + 
             T_{21} \right) .
\end{eqnarray}

The detector response can be conveniently expressed in terms of 
the complex waveform $u$:
\begin{equation}\label{detresp}
   \xi = \tilde{A} \; u + A \; \tilde{u},
\end{equation}
where $A$ and $\tilde{A}$ are the complex antenna patterns:
\begin{eqnarray}
   A & = & \frac{1}{2} \; ( F_{+} + i F_{\times} ) ,\\
   \tilde{A} & = & \frac{1}{2} \; ( F_{+} - i F_{\times} ).
\end{eqnarray}
%
A rotation $R_z(\psi)$ in the wave frame induces the 
transformation of the detector antenna patterns and the GW waveforms
\begin{eqnarray}
\label{Au}
   A'  =   e^{2i\psi} \; A ,\\
   u'  =   e^{2i\psi} \; u ,
\end{eqnarray}
but the detector response is invariant under the rotation.

\section{Likelihood analysis of gravitational wave data}
\label{overview}

In this section, we present a brief overview of 
the standard likelihood approach  to the detection
and reconstruction of gravitational wave burst signals using a network of detectors. 
Though the scientific content of this section
is essentially the same as the results in~\cite{flanagan+hughes:II,epower},
 our derivation and the notation
we use are rather different. These will aid in a clearer exposition of our 
main results in subsequent
sections. The reader is referred to~\cite{helstrom} for a textbook level 
discussion of the statistical theory of signal detection used in this paper.

\subsection{Overview}
Consider an observable that is a finite data segment ${\rm x} = \{x[1],x[2],\ldots,x[N]\}$ from a 
noisy time series. The simplest detection problem is to define a {\em decision rule }
for selecting one of two mutually 
exclusive hypotheses, $H_0$ ({\em null hypothesis}) or $H_1$ ({\em alternative hypothesis}), about the data $\bf x$. Under the $H_0$ and 
$H_1$, ${\rm x}$ is 
a realization of a stohastic process described by the joint 
probability density $p({\rm x}|H_0)$ and $p({\rm x}|H_1)$
respectively.

Any decision rule will incur two types of errors:
{\it false alarm} - $H_1$ is selected when $H_0$ is true,
and {\it false dismissal} - $H_0$ is selected when $H_1$ is true.
Each error will have a probability associated with it, namely, the false alarm 
and the false  dismissal probabilities $Q_0$ and $Q_1$ respectively.
In order to select the best decision rule, several criteria have been proposed out 
of which the Neyman-Pearson criterion is the most suitable for detection of 
gravitational waves. According to this criterion, the optimal decision rule has 
the least $Q_1$ for fixed $Q_0$. The rule
accepts $H_1$ ($H_0$) when the {\em likelihood ratio}, $\Lambda({\rm x})$, defined as
\begin{equation}
\Lambda({\rm x}) = \frac{p({\rm x}|H_1)}{p({\rm x}|H_0)}\;,
\end{equation}
is greater (less) than a threshold value that is  fixed by the specified $Q_0$.

In the case of the GW data analysis, $H_0$ is the hypothesis 
``a GW signal is absent" and $H_1$ is
``the GW signal ${\mathbf \xi}$ is present". 
For a stationary, Gaussian white noise with zero mean the corresponding joint probability densities are
\begin{eqnarray}
p({\rm x}|H_0) &=&\prod_{i=1}^{N} \frac{1}
{\sqrt{2\pi} \sigma}\exp\left(-\frac{x^2[i]}{2\sigma^2}\right)\;,\\
p({\rm x}|H_1) &=&\prod_{i=1}^{N} \frac{1}
{\sqrt{2\pi} \sigma}\exp\left(-\frac{(x[i]-\xi[i])^2}{2\sigma^2}\right)\;,
\end{eqnarray}
where $\sigma$ is the standard deviation of the noise. The logarithm of the likelihood ratio can be expressed as
\begin{equation}
{\cal L} = \ln(\Lambda({\rm x}))= \sum_{i=1}^{N} \frac{1}{\sigma^2} 
\left( x[i] \xi[i] - \frac{1}{2} \xi^2[i] \right) \;.
\end{equation}
In the rest of the paper, we will be concerned only with ${\cal L}$ which will be referred to as simply the {\it likelihood}.

The situation with two mutually exclusive hypotheses, outlined above, is the 
simplest one. In general, as in the case of GW analysis, the observed data ${\rm x}$ can be 
a realization of one among several joint probability densities $p({\rm x}|H_i)$, $i=0,1,2,\ldots$, where, as usual, $H_0$ is 
the null hypotheses and $H_i$ are the alternative hypotheses.
Correspondingly, the probabilities for false alarm and false 
dismissal can be assigned
but now the false dismissal probabilities $Q_i$ are hypothesis specific.

 One possible generalization of the Neyman-Pearson
criterion could be to select that decision rule which minimizes all 
probabilities $Q_i$ for a fixed false alarm probability $Q_0$. 
It turns out that, in general, no such rule is possible~\cite{helstrom}. 
Another approach is to generalize the 
likelihood ratio test itself by constructing 
a functional
\begin{eqnarray}
\Lambda_m({\rm x}) & = & \max_{i} \left[ \frac{p({\rm x}|H_i)}{p({\rm x}|H_0)} \right]\;,
\end{eqnarray}
and comparing it with a threshold. This test, called the maximum likelihood ratio (MLR) test,  
 tends to outperform any other {\em ad hoc} test. However, it is important to note that
the MLR test itself does not have a formal
proof of optimality. Therefore, it is possible that modifications of the MLR test, as presented in this paper,
can lead to better performance.

One of the applications of the MLR test is the detection of gravitational waves 
from the inspiral of compact binaries~\cite{finn:2001,pai:2001}. 
In principle,  the waveforms of the GW signals can be calculated 
to arbitrary precision given the parameters of the binary system.
The set of alternative hypotheses now becomes a continuum that is identified 
with the space of binary parameters. 
The likelihood ratio
$\Lambda({\rm x}|H_i)$  can, therefore, be expressed as a function over the binary parameters. 
The MLR statistic is obtained by maximizing the likelihood ratio over these parameters and reaches
its maximum for the best match of the corresponding waveform to the data.

In contrast to binary inspiral signals, where the number of parameters is small, 
the parameters characterizing burst signals
 are essentially the signal amplitudes themselves at each instant of 
time. Thus, for burst signals, the number of parameters can be very large. 
Formally, however, the concept of the likelihood ratio can still be used 
for burst signals.  In this
case, the likelihood ratio is $\Lambda({\rm x}|{\bf \xi})$, 
where ${\bf \xi}$  is the detector response to the burst signal. 
The application of the MLR test to burst signals involves maximization over each sample
$\xi[i]$ independently~\cite{mohanty:cqg04}. 

\subsection{Network likelihood}
\label{net}

So far, we have considered a time series ${\rm x}$ at the 
output of a single GW detector. 
The entire formalism outlined above can be extended to a network of detectors.
Let the data from the $k^{\rm th}$ detector be ${\rm x}_k=\{x_k[1],x_k[2],\ldots\}$ 
and the detector response to the gravitational wave be
\begin{equation}
\xi_k{[i]}=u[i]\tilde{A}_k+\tilde{u}[i]A_k. 
\end{equation}
We will assume that the noise in different detectors 
is independent. Then the likelihood ratio becomes,
\begin{eqnarray}
{\cal{L}} = \sum_{k=1}^K \sum_{i=1}^{N} \frac{1}{\sigma^2_k}
 \left( x_k[i] \xi_k[i] - 
\frac{1}{2} \xi_k^2[i]\right)\;.
\end{eqnarray}
where $K$ is the number of detectors in the network.
For detectors illuminated by the same GW source, the detector
responses are not independent. 
Therefore, the variation of the likelihood functional is performed over
the sampled amplitudes $u[i]$ and $\tilde{u}[i]$.

To characterize the angular and strain sensitivity of the network, 
we introduce the {\it network antenna patterns} 
\begin{eqnarray}
g_r = \sum_{k=1}^K{\frac{A_k\tilde{A}_k}{\sigma_k^2}}, \; \;
g_c   = \sum_{k=1}^K{\frac{A_k^2}{\sigma_k^2}},
\end{eqnarray}
where $g_r$ is real and $g_c$ is complex.
Similarly to the antenna patterns $A_k$ for a single detector, they
describe the {\it network response} to the gravitational wave:
\begin{eqnarray}
\label{response}
R(u) = g_r u + g_c \tilde{u} \;.
\end{eqnarray}
We also define the network output time series $X$ which combines the output 
time series ${\rm{x}}_k$ from individual detectors 
\begin{equation}
\label{datav}
X = \sum_{k=1}^K{\frac{{\rm{x}}_k A_k}{\sigma_k^2}}.
\end{equation}
To simplify equations, we will replace summation
over any sampled time series $s[i]$ with $\left<s\right>$.
With these new notations the likelihood functional can be written as
\begin{equation}
\label{fun1}
{\cal{L}} = 
\left< u \tilde{X} + \tilde{u} X - g_r u \tilde{u} 
- \frac{\tilde{g_c} u^2 + g_c \tilde{u}^2}{2} \right> \;,
\end{equation}
where  the $\tilde{X}$ and $\tilde{g_c}$ are complex conjugates of 
${X}$ and $g_c$ respectively.

\subsection{Solution for GW waveforms}
\label{solution}

The equations for the GW waveforms are obtained by variation of 
the likelihood functional:
\begin{eqnarray}
\frac{\delta{\cal{L}}}{\delta{u}} \; = \; 0 \;, \; \; 
\frac{\delta{\cal{L}}}{\delta{\tilde{u}}} \; = \; 0 \;,
\end{eqnarray}
which results in two linear equations for $u$ and $\tilde{u}$
\begin{eqnarray}
        X = g_r u + g_c \tilde{u}, \label{linear1}\\
\tilde{X} = g_r \tilde{u} + \tilde{g_c} u. \label{linear2}
\end{eqnarray}
The solution is
\begin{eqnarray}\label{uuu}
u_s = \frac{g_r \; X - g_c \; \tilde{X} }{g_r^2-|g_c|^2} \;.
\end{eqnarray}
Note, the solution $u_s$ satisfies the condition $X=R(u_s)$,
where $R(u_s)$ is the network response to the gravitational wave 
(see Eq.(\ref{response})).
Equations \ref{linear1} and \ref{linear2} can also be written in the matrix form
\begin{equation}
\label{eqmatrix}
   \left[ \begin{array}{c} {\mathrm{Re}(X)} \\ {\mathrm{Im}}(X) \\  \end{array} \right] =
   M_R
   \left[ \begin{array}{c} h_+ \\ h_{\times} \\  \end{array} \right] \;,
\end{equation}
where the matrix $M_R$ is given by
\begin{equation}
\label{rmatrix}
   M_R =
   \left[ \begin{array}{cc} 
            g_r+{\mathrm{Re}}(g_c)   &  {\mathrm{Im}}(g_c) \\ 
            {\mathrm{Im}}(g_c) &  g_r-{\mathrm{Re}}(g_c) \\ \end{array} \right].
\end{equation}

\subsection{Maximum likelihood ratio statistic}
\label{mlr1}

The maximum likelihood ratio statistic is obtained by
substitution of the solution $u_s$ in Eq.(\ref{fun1})
\begin{equation}\label{L0}
L_{\rm max} = \frac{2 g_r X_{r}-
\tilde{g_c} X_{c} - g_c \tilde{X}_{c}}{2 \left( g_r^2-|g_c|^2 \right)},
\end{equation}
where the quantities $X_{c}$ and $X_{r}$ are defined by
\begin{eqnarray}
X_{c} = \sum_{i=1}^K{\sum_{j=1}^K{A_i A_j D_{ij}}}, \\
X_{r} = \sum_{i=1}^K{\sum_{j=1}^K{A_i \tilde{A}_j D_{ij}}},
\end{eqnarray}
and $\tilde{X}_{c}$ is complex conjugate of $X_c$. 
The data matrix $D_{ij}$ is calculated for the detector output $\rm{x}_i$ and $\rm{x}_j$ 
scaled by the variances of the detector noise
\begin{equation}
\label{dmatrix}
D_{ij} = \frac{\left< x_i(t) x_j(t+\tau_{ij}) \right>} { \sigma^2_i \sigma^2_j} \;.
\end{equation}
The data matrix depends on the gravitational wave time delays $\tau_{ij}$ between 
the detectors. The time delays, in turn, depend on the coordinates of 
the source on the sky $\theta$ and $\phi$. 
The diagonal elements of the data matrix represent the power terms and
the non-diagonal elements represent the cross-correlation terms. 

There is a simple geometrical interpretation of the MLR statistics.
At any instance of time, the GW waveform $u$ and the network output $X$ 
can be viewed as vectors $\bf{u}$ and $\bf{X}$  
in the complex plane.
Then the MLR statistics, given by Eq.(\ref{L0}), is the inner product
\begin{equation}
L_{\rm max} = \frac{1}{2} \left< u_s \tilde{X} + \tilde{u}_s X \right> 
\; = \;
\left<\mathbf{u}_s\cdot\mathbf{X}\right> \;,
\end{equation}
which is a projection of the solution $u_s$ onto the data $X$. 
Note, that the projection is the estimator of 
the total signal-to-noise ratio of the GW signal detected in the network
\begin{equation}
{\mathrm{SNR_{tot}}} = \sum_{k=1}^K \frac{\left< \xi^2_k \right>}{\sigma^2_k} \approx 
2\left<\mathbf{u_s} \cdot \mathbf{X}\right> \;.
\end{equation}

\section{Two detector paradox}
\label{2dp}
So far we have described the standard likelihood approach for 
the detection and reconstruction of burst GW signals  wherein 
the likelihood ratio is maximized
independently over each signal sample. Though attractive because 
both a detection and estimation method are obtained simultaneously, there is a problem 
with this approach when applied to a network of two detectors. 
The problem, which we call the {\em two-detector paradox},
is described in this section. 

Let us consider a network of two
detectors in two configurations:  (A) aligned detectors and (M) misaligned detectors.
The detectors in the configuration A have the same antenna patterns. In this case
the detector responses are the same in both detectors and we consider the GW signal  
as the scalar wave $\xi$. The likelihood functional is then 
\begin{eqnarray}\label{LLA}
{\cal{L}}_A =  \frac{\left< x_1 \xi \right>}{\sigma^2_1} + 
\frac{\left< x_2\xi \right>}{\sigma^2_2}  - 
\frac{\left< \xi^2 \right>}{2} \left(  \frac{1}{\sigma^2_1} + \frac{1}{\sigma^2_2}  
\right) \; ,
\end{eqnarray}
where  $x_1$, $x_2$ are the detector outputs
and $\sigma_1$, $\sigma_2$ are the standard deviations of the detector noise. 
The solution of the likelihood variation problem is
\begin{equation}
\xi = \left( \frac{x_1}{\sigma^2_1} + \frac{x_2}{\sigma^2_2} \right) \;
      \left( \frac{1}{\sigma^2_1} + \frac{1}{\sigma^2_2} \right)^{-1}.
\end{equation}
The MLR statistics for two aligned detectors is obtained from Eq.(\ref{LLA}) 
by substituting $\xi$ with the solution
\begin{equation}
L_A = \frac{1}{2}
      \left( \frac{\left<x_1^2\right>}{\sigma^4_1} + 
      \frac{\left<x_2^2\right>}{\sigma^4_2} + 
      2 \frac{\left<x_1 x_2\right>}{\sigma^2_1 \sigma^2_2} \right) \;
      \left( \frac{1}{\sigma^2_1} + \frac{1}{\sigma^2_2} \right)^{-1}.
\end{equation}
As expected, the MLR statistic for two aligned detectors  includes
both the power and the cross-correlation terms. 
For two arbitrary misaligned detectors the MLR statistic, given by Eq.(\ref{L0}), 
reduces to
\begin{equation}
L_M = \frac{1}{2}
      \left( \frac{\left< x_1^2 \right>}{\sigma^2_1} + 
             \frac{\left< x_2^2 \right>}{\sigma^2_2} \right) \;.
\end{equation}
which includes the power terms only.

The two detector paradox is that the statistic
$L_M$ does not include cross-correlation between the detectors even for a small misalignment. 
This is highly counterintuitive since one expects that
the response of detectors to the same GW source will differ only infinitesimally when 
the detectors are infinitesimally misaligned. Hence, as in the case of $L_A$, 
one would expect that the cross-correlation term will benefit detection and 
that its importance will decline only gradually as the detectors are misaligned. 
In other words, the functional $L_M$ is expected to
approach $L_A$ in the limit of perfect alignment. 

The origin of the two detector paradox is easily seen. For the aligned case, 
the standard likelihood ratio approach has
the prior information that both detector responses are identical. 
Hence, the cross-correlation term is guaranteed to
have positive mean and, thus, should improve the detectability of GW signals. 
While, for the misaligned case, 
it is always possible to specify two arbitrary responses and 
invert them to obtain some $h_+(t)$ and $h_\times(t)$ components of the GW signal.
The standard MLR statistic, therefore, does not benefit from having 
the cross-correlation term since now it can contribute pure noise to the  statistic. 
Hence, this term disappears from the MLR statistic.  
The fact that the standard likelihood approach does not 
exhibit the expected continuity 
for the case of two detectors indicates that this approach
may not be the best one for a general network of GW detectors also.

\section{Network response}
\label{nresp}

To resolve the two detector paradox we take a closer look at
how the GW signal and the detector noise contribute to the MLR statistic.
In this section we show that the detection of 
two GW components can be considered as two independent measurements 
equally affected  by the detector noise but conducted with different angular
and strain sensitivities of the detectors. Being an {\it ad hoc} method,
the maximum likelihood may not be an optimal approach in this
situation.  For example, if the network is sensitive only to
one signal component (as in the case of co-aligned detectors)
the measurement of the second component does not benefit the
GW detection, but rather adds noise to the measurement.
In the next section we propose a solution to the problem and 
derive the detection statistics, which
continuously bridge the cases of aligned and misaligned detectors.

\subsection{Network response to gravitational waves}
\label{GWresp}

As we mentioned in Section~\ref{dresponse}, the detector response is invariant 
under rotations $R_z$ in the wave frame. Consequently, 
all measurable quantities, including the likelihood functional, 
are invariant as well.
We have a freedom to select an arbitrary wave frame by applying the
rotation $R_z(\psi)$, where $\psi$ is the rotation angle.
The rotation induces the transformation of the GW waveforms and the detector
antenna patterns (see Eq.(\ref{Au})), as well as the transformation of
the network parameters: 
$X \rightarrow X e^{i2\psi}$ and $g_c \rightarrow g_c e^{i4\psi}$. 
In general, the rotation angle $\psi$ can be selected individually for each 
instance of time and for each point in the sky.
By applying the rotation $R_z(-\gamma/4)$, where $\gamma$ is the 
phase of $g_c$, we selected a wave frame in which 
both network antenna patterns are real and positively defined. 
We call this particular coordinate frame the dominant polarization  frame . 

As follows from Eq.(\ref{response}), for a GW signal $u$ defined in the 
dominant polarization frame, the network response is 
\begin{equation}
\label{q1}
R = \left( g_r+|g_c| \right) h_{1} + 
    i \left( g_r-|g_c| \right) h_{2} \;, 
\end{equation}
where $h_{1}$ and $h_{2}$ are the real and imaginary components of the signal.
We will distinguish them from the GW polarizations $h_{+}$ and $h_{\times}$
defined for an arbitrary wave frame. Note, the coefficients in front of 
$h_{1}$ and $h_{2}$ are the eigenvalues of the network response
matrix $M_R$ (Eq.(\ref{rmatrix})), which takes a diagonal form in the 
dominant polarization frame
\begin{equation}
\label{qq1}
   M_R = g \left(
   \begin{array}{cc} 1   &  0  \\ 0 & \epsilon \\ \end{array} \right) \;.
\end{equation}
The coefficient  
\begin{equation}
\label{qqqx}
g = g_r + |g_c|
\end{equation}
characterizes
the network sensitivity to the $h_1$ wave. The sensitivity to the second 
component $h_2$ is $\epsilon g$, where $\epsilon$ is the 
network alignment factor:
\begin{equation}
\label{alfactor}
\epsilon = \frac{g_r-|g_c|}{g_r+|g_c|} \;. 
\end{equation}
The alignment factor $\epsilon$ shows the relative sensitivity of 
the network to the GW components $h_{1}$ and $h_{2}$.
Note that $0 \leq \epsilon \leq 1$. 
The total signal-to-noise ratio of the GW signal detected in the network is 
\begin{equation}
\label{q2}
{\rm SNR_{tot}} = 
2g \left( \left< h^2_{1} \right> + \epsilon \left< h^2_{2} \right> \right) \;,
\end{equation}
where $\left< h^2_{1} \right>$ and $\left< h^2_{2} \right>$ are
the sum-square energies carried by each component (see Eq.(\ref{Ess})).
Therefore, to be detected with the same 
signal-to-noise ratio, the $h_{2}$ wave should carry $1/\epsilon$ 
times more energy than the $h_{1}$ wave.

\begin{figure}[ht]
   \centering\includegraphics[width=9cm]{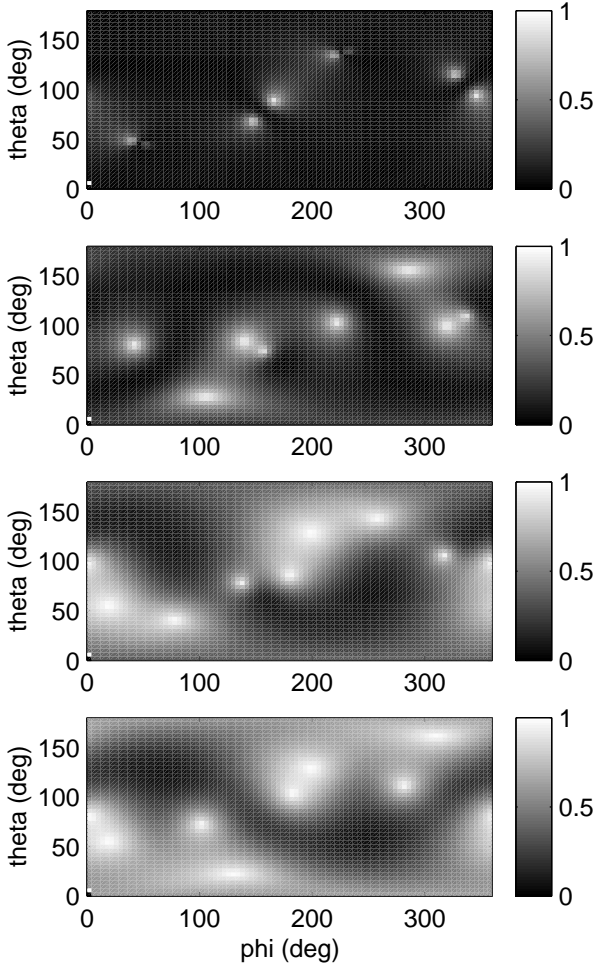}
   \caption{Alignment factors for the detector networks listed in the order from top to
            bottom: H1-L1 (upper plot), H1-L1-G1, H1-L1-G1-V1, H1-L1-G1-V1-T1 
            (bottom plot).}
   \label{factorfig}
\end{figure}

Both the network sensitivity and the alignment factor depend
on the angular and the strain sensitivities of the detectors.
The alignment factor reflects also the angular alignment of the detectors. 
For co-aligned detectors $\epsilon=0$ and the $h_2$ component of the
GW signal can not be detected. 
Even for detectors with large angular misalignment, depending on 
the sky coordinates $\theta$ and $\phi$, the alignment factor may take 
small values indicating that the detectors are effectively aligned.
For example, Figure~\ref{factorfig} shows the alignment factors
as a function of the sky coordinates calculated for several network 
configurations consisting of the H1, L1, G1, V1 and T1 detectors. 
For simplicity, we assume that the detectors have the same strain sensitivity. 
The example shows, that for the closely aligned  H1-L1 detectors, 
the alignment factor is close to zero
everywhere, except for a few small patches on the sky. The more detectors  
are added to the network, the larger is the area on the sky
with large values of $\epsilon$. 
But even for the network of five detectors (H1-L1-G1-V1-T1), 
the factor $\epsilon$ remains small for a considerable 
fraction of the sky area, where the network is 
much less sensitive to the $h_2$ wave, than to the $h_1$ wave. 
Assuming that both components carry  on average the same 
energy, the $h_2$ wave is suppressed by the factor of $\epsilon$.
Therefore, the $h_2$ component adds little to the total signal-to-noise 
ratio ${\rm SNR_{tot}}$ for GW signals originating from
areas on the sky with small values of $\epsilon$.

\begin{figure}[ht]
   \centering\includegraphics[width=9cm]{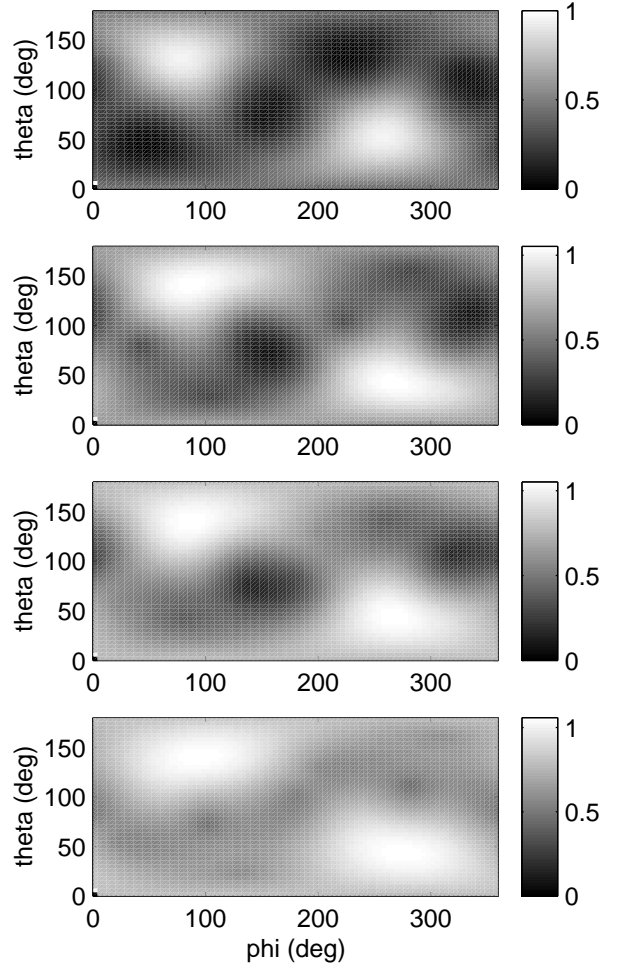}
   \caption{Sensitivity of the detector networks listed in the order from top to
            bottom: H1-L1 (upper plot), H1-L1-G1, H1-L1-G1-V1, H1-L1-G1-V1-T1 
            (bottom plot).}
   \label{sensitivity}
\end{figure}

The coefficient $g$ defines the overall sensitivity of 
the network to the gravitational waves.  
Figure~\ref{sensitivity} 
shows the network sensitivity calculated
as a function of the sky coordinates for several network 
configurations. As we expect, adding more detectors reduces the sky area 
where the network is blind to gravitational waves.

\subsection{Two components of the likelihood functional}
\label{factorization}
 
In the dominant polarization frame the likelihood functional can be written as
\begin{equation}
\label{functional}
{\cal{L}}(u) =  
\left< u \tilde{X_{\gamma}} + \tilde{u} X_{\gamma} - g_r u \tilde{u} 
- \frac{|g_c|}{2} (u^2 + \tilde{u}^2) \right> \;.
\end{equation}
where $X_{\gamma}=Xe^{-i\gamma/2}$.
Expressed in terms of $h_1$ and $h_2$, it can be written as 
${\cal{L}}(h_1,h_2)={\cal{L}}_{1}(h_1)+{\cal{L}}_{2}(h_{2})$:
\begin{eqnarray}\label{LL}
{\cal{L}}_1 = 2 \left< |X| \cos(\beta) h_1 - \frac{g}{2} h^2_1 \right> \;, \\
{\cal{L}}_2 = 2 \left< |X| \sin(\beta) h_2 - \frac{\epsilon{g}}{2} h^2_{2} \right> \;,
\end{eqnarray}
where $|X|$ is the amplitude and $\beta$ is the phase of the data vector 
$X_{\gamma}$.
The solutions for the $h_1$ and $h_2$ are obtained by the variation of 
the ${\cal{L}}_{1}$ and ${\cal{L}}_{2}$ functionals:
\begin{eqnarray}\label{q4}
h_1 = \frac{1}{g} |X| \cos(\beta) \;, 
~~~~~h_2 = \frac{1}{\epsilon{g}} |X| \sin(\beta) \;,
\end{eqnarray}
The MLR statistic can be calculated separately for each component
\begin{eqnarray}\label{q5}
L_1 = \frac{1}{g} \left< |X|^2 \cos^2(\beta) \right>=
      \frac{2 X_r + e^{-i\gamma}X_c + e^{+i\gamma}\tilde{X}_c}{4g} \;, \\
L_2 = \frac{1}{\epsilon g} \left< |X|^2 \sin^2(\beta) \right>=
      \frac{2 X_r - e^{-i\gamma}X_c - e^{+i\gamma}\tilde{X}_c}{4\epsilon g}.
\end{eqnarray}
The statistics $L_1$ and $L_2$ are the estimators of the signal-to-noise ratio 
of two GW components detected in the network.

\subsection{Detector noise}
\label{dnoise}
 
The detector output $\rm{x}_k$ is a sum of the detector noise
${n_k}$ and the detector response ${\xi_k}$. If no GW signal is present 
than the network output is
%
\begin{eqnarray}\label{eqf1}
X_n = \sum_{k=1}^K \frac{n_k A_k}{\sigma^2_k},
\end{eqnarray}
which follows from the definition of the data vector $X$ (see Eq.(\ref{datav})).
In this case, as follows from Eq.(\ref{uuu}), the likelihood 
variation procedure produces the non-zero solutions $u_s$
for the GW waveforms and the MLR statistic 
is the biased estimator of  $\rm SNR_{tot}$. 
The reconstructed sum-square energy
\begin{eqnarray}\label{eqf2}
{{E_n}} = \left<h^2_{1n}\right>+\left<h^2_{2n}\right> = 
\frac{1}{g} \left( L_{1n} + \frac{L_{2n}}{\epsilon} \right) \;,
\end{eqnarray}
is biased as well,
where $L_{1n}$ and $L_{1n}$ are the MLR statistics due to the
detector noise.
The ensemble average $\overline{{E_n}}$ can be easily calculated 
when the detector noise is white and Gaussian. Indeed, in this case 
the mean of the data matrix is  
\begin{equation}
\overline{D_{ij}} \propto \frac{\delta_{ij}}{\sigma_i \sigma_j} ,  
\end{equation}
where $\delta_{ij}$ is the Kronecker delta. 
The average MLR statistics due to the detector noise is than 
\begin{equation}
\overline{L_{1n}}=\overline{L_{2n}} \propto 1/2.
\end{equation}
As one can see, in average the detector noise introduces the same bias
for each signal component.
From Eq.(\ref{eqf2}) it follows that the reconstructed energy in the 
second component is proportional to $1/\epsilon$ and 
it diverges when $\epsilon \rightarrow 0$. 
Therefore, for small values of $\epsilon$, the likelihood variation
procedure may result in the un-physical solutions for the signal component $h_2$.
This is the root of the two detector paradox described in Section~\ref{2dp}.

The statistics $L_1$ and $L_2$ can be considered as
two independent measurements of the GW components $h_1$ and $h_2$.
Indeed, the measurements are uncorrelated, and their fluctuations are 
characterized by the same variances of the noise,
which follows from the equations 
\begin{eqnarray}
\overline{\left(L_{1n}-\overline{L_{1n}}\right) 
\left(L_{1n}-\overline{L_{1n}}\right)} = 0 ,
~~~\overline{L^2_{1n}}=\overline{L^2_{2n}} .
\end{eqnarray}
When the value of the alignment factor is small, the standard MLR
statistic $L_{\rm max}$ is not the optimal estimator of $\rm SNR_{tot}$, 
because the second component adds pure noise into the measurement.
 
\section{Constraint likelihood}
\label{constraint}
 
We have
seen that for the standard likelihood approach the problem arises
when there is a large asymmetry ($\epsilon \ll 1$) 
in the detection of two GW components. 
In this case we could find better estimators for the GW waveforms and for 
the total signal-to-noise ratio of the GW signal detected in the network. 
The construction of such estimators 
depend on our assumptions about the GW signals. Mathematically
these assumptions can be implemented as constraints applied to the likelihood
functional. The purpose of the constraints is to exclude the 
un-physical solutions arising from the variation of the likelihood 
functional. By removing such solutions
from the waveform parameter space we expect to sacrifice a small
fraction of the real GW signals, while considerably improve the 
detection for the rest of the sources. 
Below we consider examples of the likelihood constraints that can be used 
in the analysis.

\subsection{Hard constraint} 
\label{hard}

Given a source population, we could expect that in average both signal 
components $h_1$ and $h_2$ carry about the same energy. 
For example, for binary sources, the gravitational waves are 
emitted with the random inclination angles. For waves in the 
dominant polarization frame, which
is oriented randomly with respect to the source frame, the ensemble mean
of the sum-square energies of two components satisfies
\begin{equation}
\label{assh}
\overline{\left<h_1^2\right>}=\overline{\left<h_2^2\right>}.
\end{equation}
   
For areas in the sky where the network alignment factor is small, 
for most of the sources the detected energy will be dominated by the first component
(see Eq.(\ref{q2})).
For example,  for a network consisting of three interferometric detectors
H1, L1 and G1 
the alignment factor is less then 0.1 for approximately 50\% of the sky area.
Therefore, the noisy component $h_2$ can be entirely ignored for those
sky locations where $\epsilon$ is less then some threshold $\epsilon_0$.
This requirement impose a constraint on the reconstructed GW waveforms
and, therefore, on the MLR statistic.
For a given sky location we define the {\it hard} MLR statistic as
\begin{equation}
L_{\rm hard} = \left\{\begin{array}{cc}L_1 & \epsilon < \epsilon_0\;,\\
L_{\rm max} & \epsilon \geq \epsilon_0\;.\end{array} \right.
\end{equation}
When the threshold $\epsilon_0=1$, the MLR statistics is defined by 
the first signal component only. 
In the limit of a small alignment angle between the detectors 
($\epsilon \rightarrow 0$),
the hard constraint statistics converges to
the statistics for co-aligned detectors
thus resolving the two detector paradox.

The hard constraint is a good approximation in the case of
closely aligned detectors, such as the network of the H1-L1 detectors. 
The simulation results (see Section~\ref{Results}) show, that the $L_1$ 
is a reasonably good statistic
even for a network of the H1-L1-G1 detectors with large angular misalignment
between the LIGO and GEO detectors.
However, if the detection statistic $L_1$ is used, the search algorithm
is entirely inefficient to a GW signal when $h_1=0$. Although, such
GW signals are quite unlikely (due to random relative orientations of the
source and the dominant polarization frames), and for small values of $\epsilon$
they may not be detected anyway (unless the $h_2$ component is very strong),
in the next section we introduce a different constraint, which
is free from this problem.   
 
\subsection{Soft constraint} 
\label{soft}

As we mentioned in Section~\ref{factorization}, 
the unconstrained MLR statistic $L_{\rm max}$ is a sum of the statistics
$L_1$ and $L_2$, which can be written as
\begin{eqnarray}\label{soft1}
L_{\rm max} = \frac{1}{g} 
\left< |X|^2 \left(1 + \delta \right)\right>,
~~~~\delta= \frac{1-\epsilon}{\epsilon} \sin^2(\beta) \;.
\end{eqnarray}
If the detector output is dominated by the GW signal, and 
assuming that both GW components carry about the same energy,
the ensemble average for $\sin^2(\beta)$ is
\begin{equation}
\overline{\sin^2(\beta)} \approx \epsilon^2/(1+\epsilon^2), 
\end{equation}
which follows
from the expression for the network response (see Eq.(\ref{q1})). 
It means that in average 
\begin{equation}
\overline{\delta} \approx \epsilon \frac{1-\epsilon}{1+\epsilon^2} 
\end{equation}
and the second term in Eq.(\ref{soft1}) is much less then $1$. 
On contrary, for the detector noise 
\begin{equation}
\overline{\sin^2(\beta)} \approx \epsilon 
\end{equation}
and respectively
\begin{equation}
\overline{\delta} \approx 1-\epsilon.
\end{equation}
Therefore, the noisy second term in Eq.(\ref{soft1}) can be omitted, 
resulting in the statistic, which we call the {\it soft} MLR
statistic
\begin{eqnarray}\label{soft2}
L_{\rm soft} = \frac{1}{g} \left< |X|^2 \right> = L_1+\epsilon L_2  \;.
\end{eqnarray}

There is a simple statistical justification of this result.
Since the statistics $L_1$ and $L_2$ are two uncorrelated
Gaussian random variables
with the mean $\mu_{1}$ and $\mu_{2}$, and the variance $\nu$, 
the joint probability $P(L_1,L_2,\mu_1,\mu_2,\nu)$
belongs to the  Rayleigh distribution.
For the assumption above 
(see Eq.(\ref{assh})), we expect that $\mu_1=gE$
and $\mu_2=\epsilon{g}E$, where $E$ is the GW sum-square energy. 
Then the best estimator for $\rm SNR_{tot}$ 
is obtained by maximizing $P$ over $E$, which gives 
the statistic $L_{\rm soft}$. 

To obtain the solution for the GW waveforms, 
one should impose a constraint on the likelihood functional itself.
The constraint can be integrated into the variation procedure by 
the method of the Lagrange multiplier \cite{lagrange}. 
In this method, first, we have to obtain the constraint
equation. The soft constraint can be constructed by requiring that
\begin{eqnarray}\label{soft3}
g\left<h^2_{1}\right>+\epsilon g \left<h^2_{2}\right> = 0, 
- \frac{1}{g} \left< |X|^2 \right> \;.
\end{eqnarray}
which limits the sum-square energies $\left<h_1^2\right>$ and $\left<h_2^2\right>$.
Since, the constraint is applied to the $h_2$ component only, we can replace
the $h_1$ with the solution for the first component and re-write the constraint
as
\begin{eqnarray}\label{soft4}
\epsilon g \left<h^2_{2}\right> - \frac{1}{g} 
\left< |X|^2 \sin^2(\beta)\right> = 0 \;.
\end{eqnarray}
The solution for the second GW component $h_{\rm 2}$ is trivially obtained by the 
constraint variation of the likelihood functional ${\cal{L}}_2$
\begin{eqnarray}\label{soft5}
h_{\rm 2soft} = \frac{1}{\sqrt{\epsilon}g} |X| \sin(\beta) \;.
\end{eqnarray}
As one can see, the constrained solution is the standard solution $h_2$, 
multiplied by a penalty factor of $\sqrt{\epsilon}$,
which reduces both the noise and the signal contribution from 
the second component to the MLR statistic at small values of ${\epsilon}$. 
Obviously, it reduces the sensitivity of the $L_{soft}$ statistic 
to a particular class of GW signals with $h_1=0$, described in Section~\ref{hard}.
For these signals, to be detected with the same false alarm rate, 
the $L_{\rm soft}$ statistic requires 
$(1+\epsilon^2)/2\epsilon$ times more powerful GW signal,
then the standard $L_{\rm max}$ statistic.  
For example, for $\epsilon=0.1$ the degradation of the strain 
sensitivity is by a factor of 2. But it happens only for a small fraction
of the GW sources. Compare to the standard likelihood method, 
for most of the sources we expect to improve the 
detection sensitivity if the $L_{\rm soft}$ statistic is used.


\begin{figure}[t]
   \centering\includegraphics[width=0.45\textwidth]{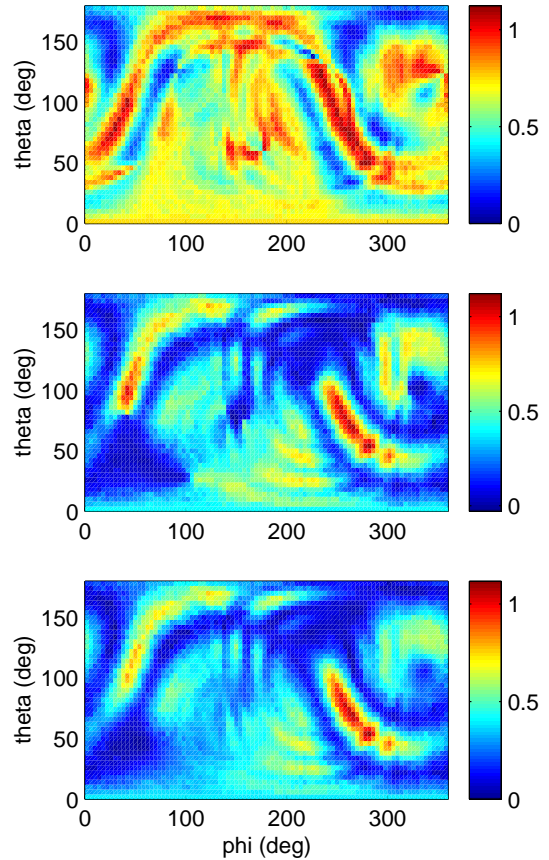}
   \caption{Sky maps of the likelihood statistics $L_{\rm max}$ (top),
   $L_{\rm hard}$ (middle) and $L_{\rm soft}$ (bottom)  for the detector
   network H1-L1-G1. The injected signal SNR is $17$ (H1),  $20$
   (L1) and $9$ (G1). The source is located at $\theta = 50^{\circ}$ and
   $\phi = 280^{\circ}$.}
   \label{3DetSkyMaps}
\end{figure}

\subsection{Network sky maps}

In un-triggered burst searches the coordinates of the
source, $\theta$ and $\phi$ are free parameters. 
In this case, the
detector responses, the likelihood statistics and the reconstructed
waveforms become functions of $\theta$ and $\phi$ or skymaps.
For example, the skymaps corresponding to different  statistics are 
shown in Figure~\ref{3DetSkyMaps}. (For details see Section~\ref{Results}).

For a given location in the sky, the value of the likelihood statistic 
indicates how consistent the data
 is with the hypothesis that a GW signal originates 
from that location. The coordinates $\theta$ 
and $\phi$ which yield maximum for the likelihood statistic correspond
to the most probable location of the source. 
The maximum value of the statistic is then used for detection. 
By setting a threshold on the 
maximum likelihood value one can decide on the
presence or absence of a gravitational wave signal in the data as described in 
Section~\ref{overview}. Given the most probable source coordinates 
the waveforms $h_+(t)$ 
and $h_{\times}(t)$ are reconstructed as described in Section~\ref{constraint}.

\section{Numerical Simulation}
\label{Results}

We have outlined a general method for using the MLR statistic in
conjunction with constraints limiting the space of the GW
waveforms. The method is intended for application to burst searches
with networks of gravitational wave detectors. The performance of 
the method and the effect of the constraints can be analyzed using 
numerical simulations with modeled waveforms. The present  
simulation is  similar to the one previously used for
estimating the performance of the mixed correlation method
\cite{rakhmanov+klimenko}.

\subsection{Simulation procedure}

The two-polarization waveforms which represent burst gravitational
waves used in the simulation are taken from the numerical models of the 
merger phase of coalescing binary
black holes (BH)~\cite{lazarus}. These waveforms form a one-parameter
family BH-$M$, where $M$ is the total mass of the binary system
in units of solar mass. The results below correspond to $M=100$. 
In the simulations
we generate the detector noise which is Gaussian and white. The variance of the
noise is selected to be the same for all detectors.

A typical simulated data segment has the duration of 1 second and
consists of $N=4096$ data samples.
For calculation of the data matrix (see Eq.(\ref{dmatrix})) we set 
the integration window of 85~ms, which is substantially greater than the 
duration of the signal. The magnitude of 
the simulated signals is controlled by the overall gain $G$, which is 
varied from $0$ to $10$, whereas the magnitude of the noise is kept fixed.

Due to different orientation of the detectors with respect to the
incoming gravitational wave, the detectors  
responses  are different (see Eq.(\ref{detresp})). 
To characterize the magnitude of the signal in any
given detector we define the signal-to-noise ratio: 
\begin{equation}
   {\mathrm{SNR}} = \int_{-\infty}^{\infty} 
      \frac{|{\xi}(f)|^2}{S(f)} df 
   \rightarrow \frac{1}{\sigma^2} 
      \sum_{i=0}^{N-1} \xi^2(t_i)  ,
\end{equation}
where $S(f)$ is the power spectral density of the noise. For white
Gaussian noise $S(f) = \sigma^2/f_s$, where $f_s$ is the sampling
rate.

For any given detector in the network, the magnitude of the signal 
varies significantly depending on the source location in the sky. 
We therefore choose the location of the simulated sources at random,  
with a uniform distribution over the sky. We also choose the
polarization angle $\psi$ at random from the interval 
$[0^{\circ}, 360^{\circ}]$. With these choices the simulation gives us
an estimate of the performance of the detection algorithms without the 
bias which can be introduced by the particular choice for the 
source location or its polarization angle.

The simulation consists of series of tests corresponding to 
different values of SNR (controlled by $G$). For each value of $G$
a total number of 10,000 injections were made. 
To characterize the strength of the signal in each detector for the
entire test we introduce the sky-average SNR, denoted by 
$\overline{\mathrm{SNR}}$. The sky-average SNR is proportional to $G$ 
and it is the same for each detector in the network 
($\overline{\mathrm{SNR}} \approx 2.3 G$).

\subsection{Simulation results}

\begin{figure*}[t]
   \includegraphics[width=0.45\textwidth]{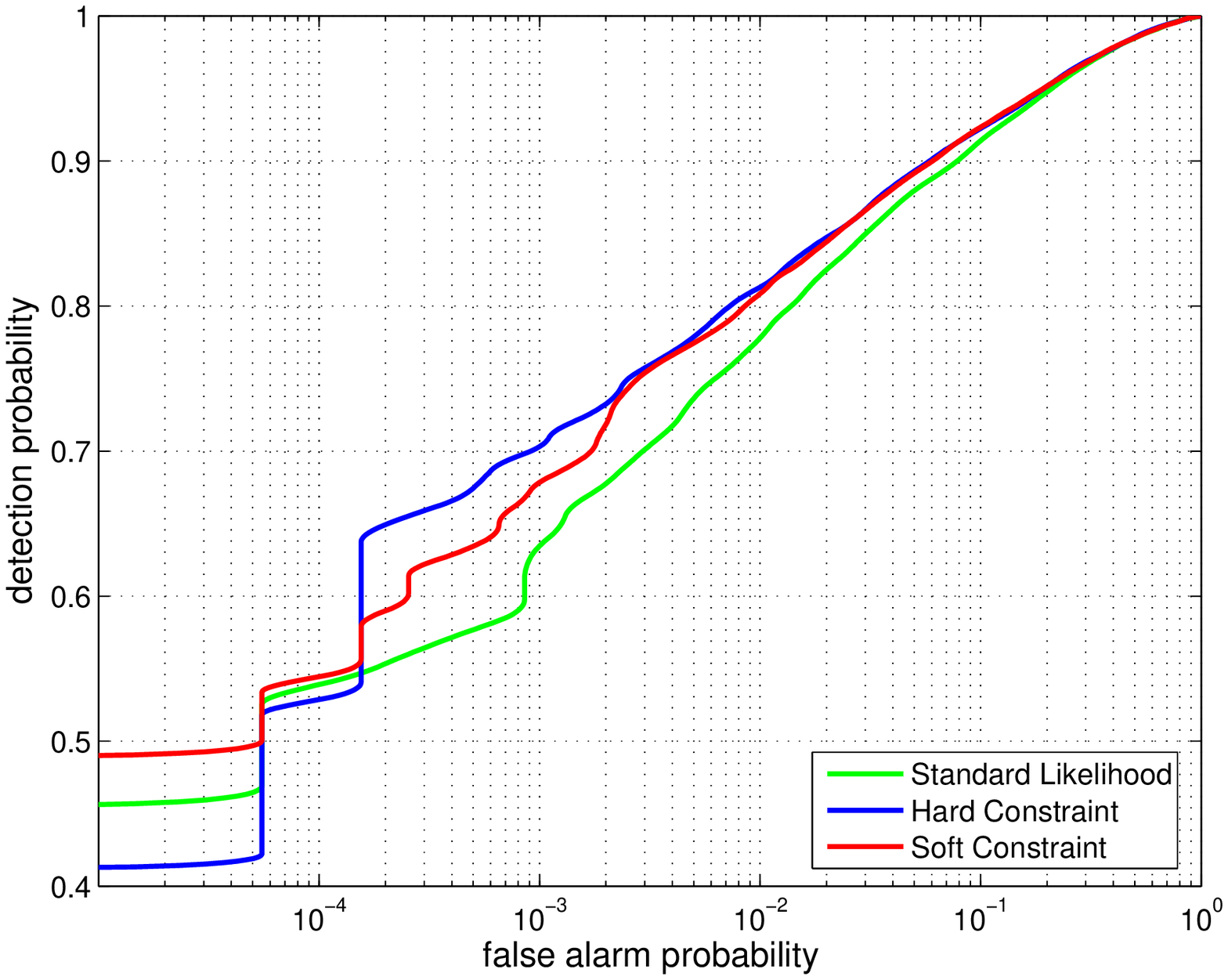}
   \includegraphics[width=0.45\textwidth]{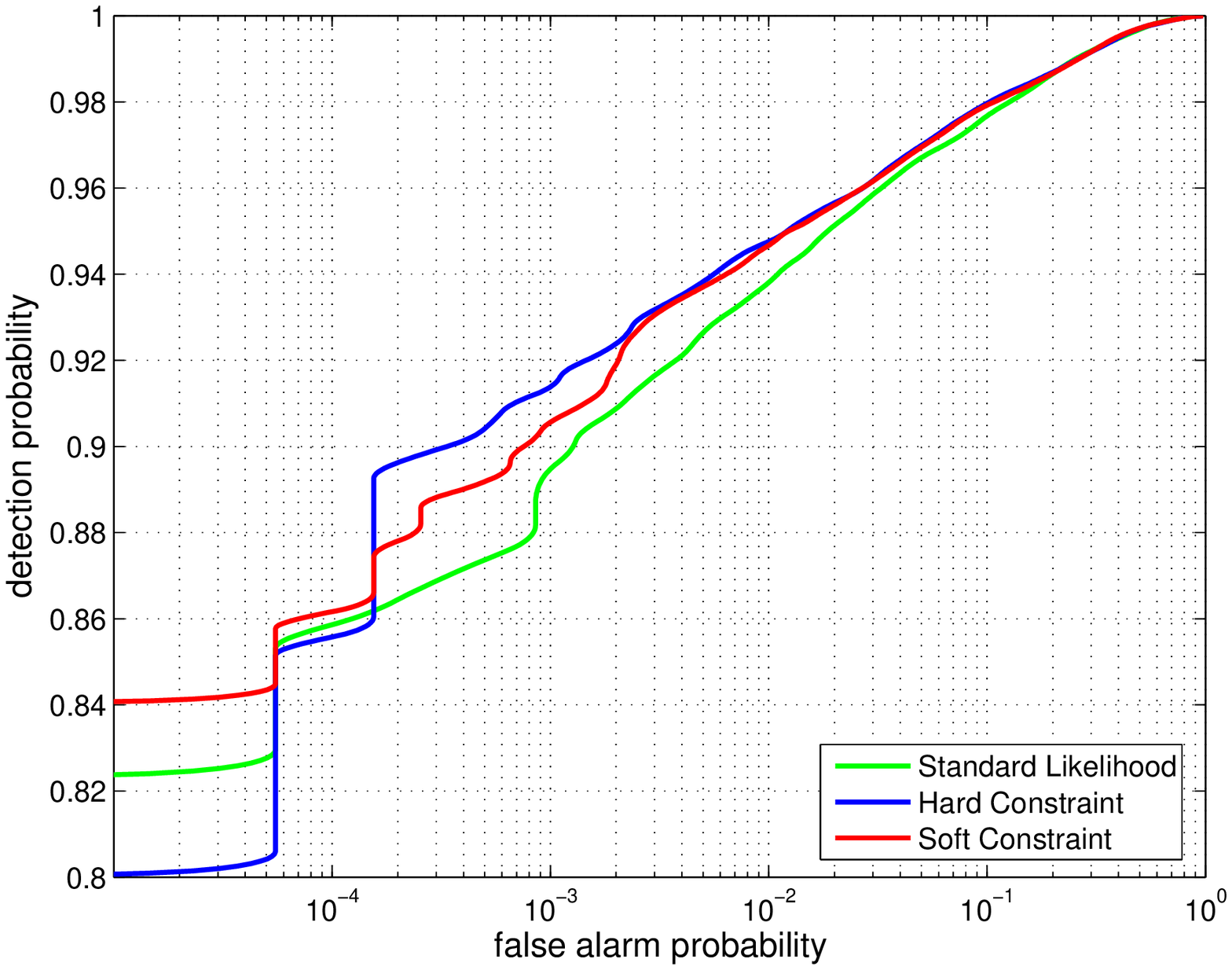}
   \caption{Receiver operation characteristics for the network of H1-L1-G1
   detectors: $\overline{\mathrm{SNR}} = 6.9$ (left) and 
   $\overline{\mathrm{SNR}} = 9.2$ (right).}
   \label{rocDetS3}
\end{figure*}

In the simulation we tested the following detection methods:
the standard likelihood method 
($L_{\mathrm{max}}$), the hard constraint method 
($L_{\mathrm{hard}}$), 
and the soft constraint nethod ($L_{\mathrm{soft}}$). 
The detection performance of the methods  is compared
by using the receiver operating characteristic (ROC),
which shows the detection probability as a function of the 
false alarm probability. 
Examples of the ROC curves, corresponding to $G=3$ and
$G=4$, are shown in Figure~\ref{rocDetS3}.

\begin{figure}[t]
  \centering\includegraphics[width=0.45\textwidth]{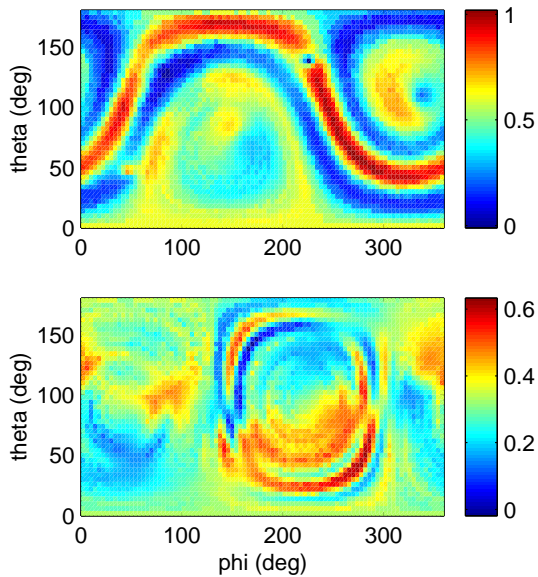}
   \caption{Sky maps of network statistics for 2 detector networks H1-L1 (top)
             and H1-G1 (bottom). The source is located at $\theta = 50^{\circ}$
             and $\phi = 280^{\circ}$.}
   \label{2DetSkyMaps}
\end{figure}

The accuracy of the source localization depends 
on the strength of the GW signal and on the network
configuration. With only one detector in the network, the likelihood 
statistics is constant across the sky (it has no $\theta$ or $\phi$
dependence) and therefore the source localization is not possible. 
However, already with two spatially separated
detectors 
the network becomes sensitive to the source location (see Figure~\ref{2DetSkyMaps}).
In the case of two closely aligned detectors H1-L1, 
the area with the large values of the likelihood is rather a ring then a point, 
showing an ambiguity in the determination of the source location. 
But even in this case the method gives directional information about the source 
and allows exclusion of the most of the sky area as inconsistent with the detected 
GW signal. For two misaligned detectors H1-G1, the source localization is more 
accurate due to different angular sensitivities of the detectors.
Even more accurate estimation of the source coordinates can be obtained with
three and more detectors in the network (see Figure~\ref{3DetSkyMaps}). 
The greater the number of the detectors 
in the network, the better the source localization. 



The error in the source localization is given 
by the angle $\alpha$ between the true direction to the source and 
the reconstructed direction to the source. Equivalently, $\alpha$ can
be defined as the length of an arc connecting these two locations on a
sphere with unit radius. 
To describe the efficiency of the source localization we introduce the
following figure of merit. First, we chose a cone with 
the opening angle $\alpha_c$ which constitutes an acceptable error.
Then we calculate the number of detected sources ($N_\alpha$) which satisfy
the condition $\alpha < \alpha_c$.
The ratio of $N_\alpha$ to the total number of injections 
defines the efficiency of the source localization
and depends on the signal-to-noise ratio $\overline{SNR}$.

\begin{figure*}[t]
   \includegraphics[width=0.45\textwidth]{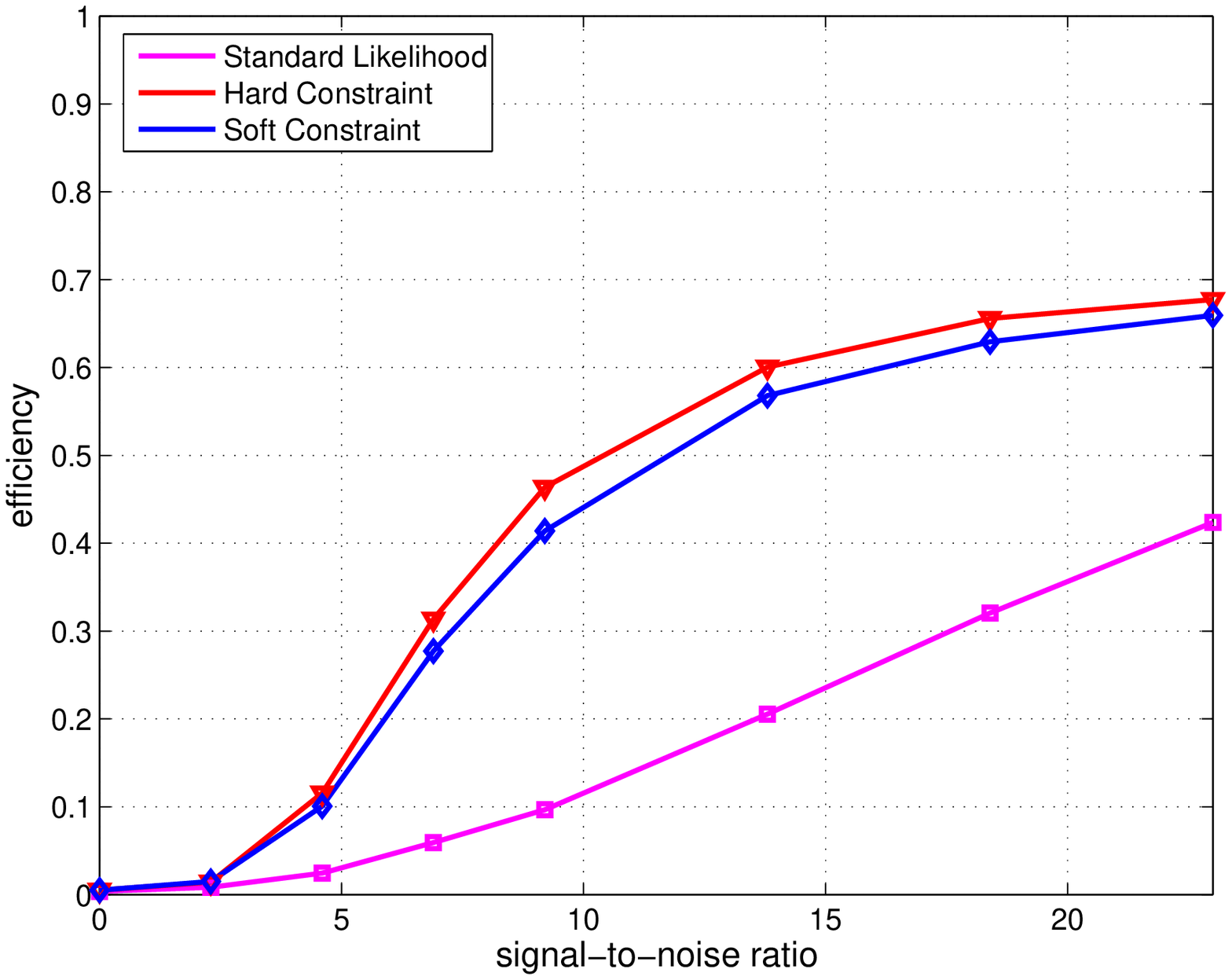}
   \includegraphics[width=0.45\textwidth]{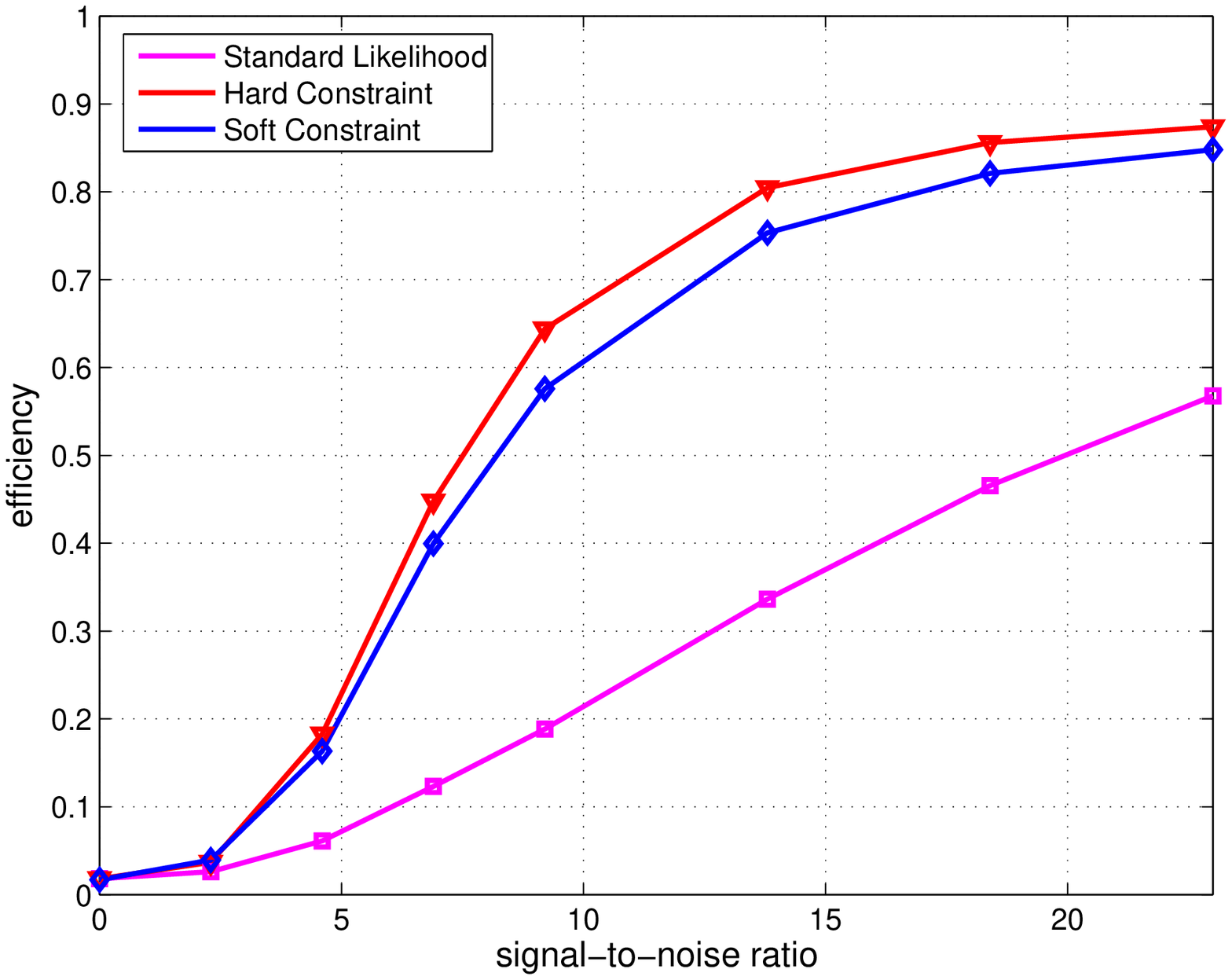}
   \caption{Efficiency of the source localization with a network of
   three detectors H1-L1-G1: $\alpha_c = 8$ (left), $\alpha_c = 16$ (right).}
   \label{angErr}
\end{figure*}

Figure~\ref{angErr} shows the efficiencies of the source localization
for the L1-H1-G1 network corresponding to the different detection methods. 
In this example, 
the values of the acceptable localization error are chosen to be 
$\alpha = 8^{\circ}$ and $\alpha = 16^{\circ}$. 
Note that the constraint likelihood methods
perform considerably better than the standard
likelihood method. Let us consider, 
for example, the source localization for $\overline{\mathrm{SNR}}$ of 10 (20) .
The hard constraint method recovers 
approximately $48\%$ ($66\%$) of all simulated sources within the 
$8$-degree angle from their true location. In comparison, 
the standard likelihood method  yields only 
$12\%$ ($35\%$) efficiency for the same angle. 
Within the $16$-degree angle, the hard constraint method 
 recovers $66\%$ ($86\%$) of all the simulated sources,  whereas 
the standard method yields only $22\%$ ($50\%$) efficiency. 
Similar comparisons hold for the soft 
constraint method. We find that
for both constraint likelihood methods, the events with 
poorly reconstructed coordinates come from the areas in the sky with 
small values of the network sensitivity.


\section{Conclusion}

We have presented a novel approach to the detection and reconstruction of gravitational 
waves with an arbitrary network of interferometric detectors. Starting with 
the network likelihood ratio functional for unknown gravitational wave burst signals, 
we identify and solve the two detector paradox. The essence of the paradox is that 
in the case of two arbitrary misaligned detectors the maximum
likelihood ratio statistics depends only on the power in the
detector data streams. It does not agree with the statistic of two 
co-aligned detectors, which depends also on the cross-correlation between
the detectors. We show that the problem is associated with the different
sensitivity of the detector network to two polarization components of the GW
signal and present not only in the case of two detectors, but for any
arbitrary network. To characterize the difference in
the sensitivity to the GW components, we introduce the network alignment factor.
For locations on the sky where the value of the alignment factor is small, the network
is sensitive to only one GW component and the variation of the likelihood
functional results in the un-physical solutions for the second GW component.
To exclude the un-physical solutions we propose to use constraints, which
limit the parameter space of the GW waveforms  and result in a new class of the
maximum likelihood ratio statistics. For the networks of two and more detectors,
the constraint likelihood methods allow reconstruction of the two GW polarization
components and the location of the source on the sky. 

In the paper we introduce two examples
of the constraint statistics, which performance is compared with 
the standard likelihood statistics.
The performance of the method was estimated with the numerical simulation.
We restricted our simulation to the case of the white Gaussian noise. 
For simplisity we assumed that all detectors have identical sensitivities though the 
method presented in this paper does not have these restrictions. 
Our simulation results indicate that the constraint likelihood method 
enhance the detection of the GW signals and performs 
significantly better than the standard likelihood method in the 
reconstruction of the source coordinates. 
We believe that since all the methods we have considered are compared on exactly the same
footing, our results regarding relative performance will not change for the general case. 
However, as a work in progress, we plan to expand our simulations to more realistic 
detector noise.

\section{Acknowledgments}
We thank Jolien Creighton for usefull discussion and comments on the paper.
We also thank Erik Katsavounidis, Peter Saulson, Massimo Tinto and Patrik Sutton 
for comments on the paper.
This work was supported by the US National Science Foundation grants PHY-0244902, PHY-0070854 to the University of Florida, Gainesville and
NASA grant NAG5-13396 to the Center for Gravitational Wave Astronomy at the University of Texas at Brownsville.

\end{document}